\documentclass[12pt]{article}
\usepackage[T1]{fontenc}

\usepackage[a4paper,centering,body={15.27cm,22.50cm}]{geometry}

\usepackage[numbers,sort]{natbib}
\usepackage{graphicx}
\usepackage{amsmath}
\usepackage{amsfonts}
\usepackage{amssymb}
\usepackage{amsthm}
\title{About several classes of bi-orthogonal polynomials and discrete integrable systems}
\author{Xiang-Ke Chang\footnotemark[1] \footnotemark[2] , Xiao-Min Chen\footnotemark[1] \footnotemark[2], Xing-Biao Hu\footnotemark[1] and Hon-Wah Tam\footnotemark[3]}

\begin{document}
\date{}
\maketitle

\footnotemark[1]{LSEC, Institute of Computational Mathematics and
  Scientific Engineering Computing, AMSS, Chinese Academy of Sciences,
  P.O.Box 2719, Beijing 100190, PR China}
\footnotemark[2]{University of Chinese Academy of Sciences, Beijing, PR China}
\footnotemark[3]{Department of Computer Science, Hong Kong Baptist University, Kowloon Tong, Hong Kong, China}
\let\thefootnote\relax\footnotetext{Emails: changxk@lsec.cc.ac.cn,chenxm@lsec.cc.ac.cn,hxb@lsec.cc.ac.cn,tam@comp.hkbu.edu.hk}
\begin{abstract}
By introducing some special bi-orthogonal polynomials, we derive the
so-called discrete hungry quotient-difference (dhQD) algorithm and a
system related to the QD-type discrete hungry Lotka-Volterra
(QD-type dhLV) system, together with their Lax pairs. These two
known equations can be regarded as extensions of the QD algorithm.
When this idea is applied to a higher analogue of the discrete-time
Toda (HADT) equation and the quotient-quotient-difference
(QQD) scheme proposed by Spicer, Nijhoff and van der Kamp, two extended systems are constructed. We call these systems the hungry forms of the higher analogue discrete-time Toda
(hHADT) equation and the quotient-quotient-difference
(hQQD) scheme, respectively. In addition, the corresponding Lax pairs are
provided.
\end{abstract}


\vspace{2pc}
\maketitle

\section{Introduction}
  Discrete integrable systems richly connect many areas of mathematical physics and other fields, e.g., orthogonal polynomials,  numerical algorithms, and combinatorics. This paper is devoted to the research between discrete integrable systems and orthogonal polynomials.

  There has been growing interest in studying the relationship between discrete integrable systems and orthogonal polynomials, as orthogonal polynomials provide a powerful tool for studying discrete integrable systems \cite{Chudnovsky1983laws,spiridonov1995discrete}. For example, in \cite{papageorgiou1995orthogonal}, the quotient-difference (QD) algorithm \cite{rutishauser1954quotienten}
    \begin{equation}\label{qd}
  \begin{array}{l}
  e_k^{l+1}q_k^{l+1}=e_k^{l}q_{k+1}^{l} \; ,\\
  e_{k}^{l+1}+q_{k+1}^{l+1}=e_{k+1}^{l}+q_{k+1}^{l}
  \end{array}
  \end{equation}
  is nothing but the compatibility condition of the spectral problem related to the discrete-time Toda equation \cite{hirota1977nonlinear}
  \begin{equation*}
  \tau_{k+1}^{l-1}\tau_{k-1}^{l+1}=\tau_{k}^{l-1}\tau_{k}^{l+1}-(\tau_{k}^{l})^2.
  \end{equation*}
   (We remark that, throughout this paper, the discrete indices are integers except otherwise specified.) The compatibility of a special type of symmetric orthogonal polynomials yields the discrete-time Lotka-Volterra chain \cite{spiridonov1997discrete}.  The (2+1)-dimensional Toda lattice can be obtained by using formal bi-orthogonal polynomials \cite{tsujimoto2000discrete}. More examples can be found in references such as \cite{Miki2011cauchy,spiridonov1995discrete,spiridonov2000spectral,spiridonov2007integrable,tsujimoto2000discrete,vinet1998integrable}.
   The following table, which is by no means exhaustive, summarizes the relationships stated in the above references:
      \begin{table}[h] \small
 \centering
\caption{Some examples relating orthogonal polynomials and discrete integrable systems}
\begin{tabular}{|c|c|c|}
  \hline
  Orthogonal polynomials (OP) & Discrete integrable systems (dIS) &Ref.\\ \hline 
  standard OP & discrete-time Toda equation&\cite{papageorgiou1995orthogonal,spiridonov1995discrete}\\ \hline
  symmetric OP & discrete-time Lotka-Volterra chain&\cite{spiridonov1997discrete}\\ \hline
  string OP&   (2+1)-dimensional Toda lattice&\cite{tsujimoto2000discrete}\\ \hline
  a class of OP & discrete-time hungry Lotka-Volterra chain&\cite{tsujimoto2000discrete}\\ \hline
 $ R-\uppercase \expandafter {\romannumeral 1}$ polynomials &a generalized relativistic Toda chain&\cite{vinet1998integrable}\\ \hline
  $ R-\uppercase \expandafter {\romannumeral 2}$ polynomials & $ R-\uppercase \expandafter {\romannumeral 2}$ chain&\cite{spiridonov2000spectral}\\ \hline
  FST polynomials& FST chain&\cite{spiridonov2007integrable}\\ \hline
  Cauchy bi-orthogonal polynomials & a special dIS&\cite{Miki2011cauchy}\\ \hline
  quasi-orthogonal polynomials& another special dIS&\cite{Miki2011cauchy}\\ \hline
  a class of two-variable OP& higher order analogue of discrete-time Toda&\cite{spicer2011higher}\\
\hline
\end{tabular}
\end{table}

  Based on these observations, we plan to enrich the connections between orthogonal polynomials and discrete integrable systems.

  In this paper, we first derive two known discrete integrable systems, namely, the so-called discrete hungry quotient-difference (dhQD) algorithm \cite{tokihiro1999proof} and a system \cite{fukuda2011backlund} related to the QD-type discrete hungry Lotka-Volterra (QD-type dhLV) system, by introducing a novel bi-orthogonal condition. The corresponding bi-orthogonal polynomials satisfy a linear recurrence relation. To the best of our knowledge, this is the first time that connections are derived between these two systems and orthogonal polynomials.
  The details will be presented in Section 2.

  Additionally, in a recent paper \cite{spicer2011higher}, Spicer et al. have proposed a higher analogue of the discrete-time Toda (HADT) equation
  \begin{equation}\label{HADT}
  \begin{array}{l}
  \sigma_{k+1}^{l-2}(\sigma_{k-1}^{l+1}\sigma_{k}^{l+2}\sigma_{k}^{l-1}-\sigma_{k-1}^{l+1}\sigma_{k}^{l}\sigma_{k}^{l+1}+\sigma_{k-2}^{l+2}\sigma_{k}^{l+1}\sigma_{k+1}^{l-1})\\
  =\sigma_{k-1}^{l+2}(\sigma_{k}^{l-1}\sigma_{k-1}^{l+1}\sigma_{k+2}^{l-2}-\sigma_{k}^{l-1}\sigma_{k}^{l}\sigma_{k+1}^{l-1}+\sigma_{k}^{l+1}\sigma_{k+1}^{l-1}\sigma_{k}^{l-2})
  \end{array}
  \end{equation}
  and a so-called quotient-quotient-difference (QQD) scheme
   \begin{equation}\label{qqd}
  \begin{array}{l}
  u_k^{l+3}w_k^{l+1}=u_{k+1}^{l+1}w_{k+1}^{l+1} \; ,\\
  u_k^{l+3}v_k^{l+1}=u_{k+1}^lv_{k+1}^l \; ,\\
  u_{k}^{l+3}+v_{k+1}^{l+1}+w_{k+1}^l=u_{k+2}^l+v_{k+1}^l+w_{k+1}^{l+1}
  \end{array}
  \end{equation}
  by considering two-variable orthogonal polynomials restricted to an elliptic curve. Our second objective is to give generalizations to these two systems, calling them the hungry forms of the HADT (hHADT) equation and the QQD (hQQD) scheme, respectively, by constructing orthogonal conditions. This part will be described in Section 3.

  In the literature, the recurrence relations of many algorithms in numerical analysis can be regarded as integrable systems. For example, one step of the QR algorithm is equivalent to the time evolution of the finite nonperiodic Toda lattice \cite{symes1982qr}. Wynn's celebrated $\varepsilon$-algorithm \cite{wynn1956device} is nothing but the fully discrete potential KdV equation \cite{nakamura2000applied,papageorgiou1993integrable}. On the other hand, some discrete integrable systems have been used to design numerical algorithms. For instance, the discrete Lotka-Volterra equation can be used as an efficient algorithm to compute singular values \cite{iwasaki2002convergence,iwasaki2004application,tsujimoto2001discrete}. Several convergence acceleration algorithms have been derived by virtue of discrete integrable systems \cite{brezinski2010multistep,he2011convergence,sun2013extended}.
  Additionally, since the dhQD algorithm can be used to compute eigenvalues of a totally nonnegative banded matrix \cite{fukuda2012integrable}, we hope also that the hQQD scheme has future applications in numerical algorithms.

\section{The dhQD algorithm and the system related to QD-type dhLV via an orthogonal condition}
  In this section, we derive the dhQD algorithm \cite{tokihiro1999proof} and a system \cite{fukuda2011backlund} related to the QD-type dhLV equation using some special bi-orthogonal polynomials. Both of these two processes can be reduced to 
  the ordinary orthogonal polynomials case, which yields the discrete-time Toda equation.

  For a fixed positive integer $m$, let $P_n^l(m,x)$, $Q_n^l(m,x)$, $n=0,1,2,...,l=0,1,2,...$ be two classes of adjacent monic polynomials satisfying the bi-orthogonal condition
  \[
  <P_k^l(m,x),Q_n^l(m,x)>_m^l=h_k^l(m)\delta_{kn},
  \]
  where $<\cdot,\cdot>_m^l$ is a bilinear form defined by $<f(x),g(x)>_m^l=\int_{\Gamma} f(x^m)g(x)w(x)x^ldx$. In general, both the weight function $w(x)$ and the contour of integration $\Gamma$ are complex. In the sequel, we omit the index $m$ for abbreviation.

  If we define the moments $c_i=\int_{\Gamma} w(x)x^idx$, it is easy to check that $P_n^l(x)$ and $Q_n^l(x)$ may be expressed as
\[
P_n^l(x)=\frac{1}{\tau_n^l}\left|\begin{array}{ccccc}
  c_l&c_{l+1}&\cdots&c_{l+n-1}&1\\
  c_{l+m}&c_{l+m+1}&\cdots&c_{l+m+n-1}&x\\
  \vdots&\vdots&\ddots&\vdots&\vdots\\
  c_{l+mn}&c_{l+mn+1}&\cdots&c_{l+mn+n-1}&x^n\\
  \end{array}
\right|
\]
and
\[
Q_n^l(x)=\frac{1}{\tau_n^l}\left|\begin{array}{ccccc}
  c_l&c_{l+1}&\cdots&c_{l+n}\\
  c_{l+m}&c_{l+m+1}&\cdots&c_{l+m+n}\\
  \vdots&\vdots&\ddots&\vdots\\
  c_{l+m(n-1)}&c_{l+m(n-1)+1}&\cdots&c_{l+m(n-1)+n}\\
  1&x&\cdots&x^n
  \end{array}
\right|,
\]
where $\tau_n^l$ denotes $\det(c_{l+mi+j})_{i,j=0}^{n-1}$ and we use the convention that $\tau_0^l=P_0^l=Q_0^l=1$ and $P_{-1}^l=Q_{-1}^l$=0. We remark that $P_n^l$ and $Q_n^l$ both satisfy $m+2$ term recurrence relations. In fact, when one expands $xP_n^l$ and $x^mQ_{n-m+1}^l$ as
\begin{eqnarray*}
xP_n^l&=&P_{n+1}^l+a_{n,n}^lP_n^l+a_{n,n-1}^lP_{n-1}^l+\cdots+a_{n,0}^l,\\
x^mQ_{n-m+1}^l&=&Q_{n+1}^l+b_{n,n}^lQ_n^l+b_{n,n-1}^lQ_{n-1}^l+\cdots+b_{n,0}^l,
\end{eqnarray*}
one obtains
\[
a_{n,i}^l=b_{n,i}^l=0,\ \ i=0,1,\cdots,n-m-1
\]
by orthogonality. Thus $m+2$ term recurrence relations are satisfied.

  Note also that the $P_n^l$ and $Q_n^l$ studied in this section are both particular cases of adjacent families of formal bi-orthogonal polynomials \cite{brezinski1992biorthogonality,brezinski2002computational} (or string orthogonal polynomials \cite{adler1997string,tsujimoto2000discrete}).
  In addition, $m$ is fixed for all values of $n$ in our $P_n^l$ and $Q_n^l$, whereas $m$ varies with $n$ for general bi-orthogonal polynomials.

  Now consider relations among $P_n^l$.
  Employing the two-row/column Sylvester identity (or Jacobi identity \cite{aitken1959determinants,brualdi1983determinantal,spicer2011higher}) to the determinants
    \[D_1=\left|\begin{array}{ccccc}
  1&c_l&c_{l+1}&\cdots&c_{l+n-1}\\
  x&c_{l+m}&c_{l+m+1}&\cdots&c_{l+m+n-1}\\
  \vdots&\vdots&\ddots&\vdots&\vdots\\
  x^n&c_{l+mn}&c_{l+mn+1}&\cdots&c_{l+mn+n-1}\\
  \end{array}
\right|
  \]
  and
  \[D_2=\left|\begin{array}{ccccc}
    0&0&\cdots&1&0\\
  c_l&c_{l+1}&\cdots&c_{l+n-1}&1\\
  c_{l+m}&c_{l+m+1}&\cdots&c_{l+m+n-1}&x\\
  \vdots&\vdots&\ddots&\vdots&\vdots\\
  c_{l+m(n-1)}&c_{l+m(n-1)+1}&\cdots&c_{l+m(n-1)+n-1}&x^{n-1}
  \end{array}
\right| \; ,
  \]
  we have
 \begin{eqnarray*}
D_i D_i\left(\begin{array}{cc}
1 & n+1 \\
1 & n+1 \end{array}\right)&=&D_i\left(\begin{array}{c}
1  \\
1 \end{array}\right)D_i\left(\begin{array}{c}
n+1  \\
n+1 \end{array}\right)-D_i\left(\begin{array}{c}
1  \\
n+1 \end{array}\right)D_i\left(\begin{array}{c}
n+1  \\
1 \end{array}\right), \; i=1,2,
\end{eqnarray*}
where $D\left(\begin{array}{cccc}
i_1&i_2 &\cdots& i_k\\
j_1&j_2 &\cdots& j_k
\end{array}\right)$ denotes the
determinant of the matrix obtained from a matrix $D$ by removing the rows with indices
$i_1,i_2 ,\cdots, i_k$ and the columns with indices $j_1,j_2,\cdots j_k$.
These two relations are equivalent to
\begin{eqnarray}
P_{n}^{l+m}&=&\frac{1}{x}(P_{n+1}^l+v_n^lP_n^l) \; ,\label{dhRLV_p1}\\
P_{n+1}^{l}&=&P_{n+1}^{l+1}+w_{n}^lP_{n}^{l+1}\label{dhRLV_p2} \;,
\end{eqnarray}
with
\[
v_n^l=\frac{\tau_{n+1}^{l+m}\tau_{n}^{l}}{\tau_{n}^{l+m}\tau_{n+1}^{l}},\ \ \ \
w_n^l=\frac{\tau_{n+2}^{l}\tau_{n}^{l+1}}{\tau_{n+1}^{l+1}\tau_{n+1}^{l}}.
\]
The compatibility of relations (\ref{dhRLV_p1}) and (\ref{dhRLV_p2}) leads to  a system \cite{fukuda2011backlund} related to the QD-type dhLV equation:
  \begin{eqnarray}
  w_n^{l+m}v_n^{l+1}=w_n^{l}v_{n+1}^{l} \;,\label{dhRLV1}\\
  w_{n-1}^{l+m}+v_n^{l+1}=w_n^{l}+v_n^{l} \;.\label{dhRLV2}
  \end{eqnarray}
With the help of (\ref{dhRLV_p1}) and (\ref{dhRLV_p2}), we can also obtain a Lax pair for (\ref{dhRLV1}) and (\ref{dhRLV2}). Let
\begin{equation*}
\Psi_n^l=(P_n^{l+m},P_n^{l+m-1},\cdots,P_n^{l})^T,
\end{equation*}
then
\begin{equation}\label{dhrlv1-lax}
\Psi_{n+1}^l=L_{n}^{l}\Psi_{n}^l \;,\ \ \ \Psi_{n}^{l+1}=M_{n}^{l}\Psi_{n}^l \;,
\end{equation}
with
\[
L_{n}^{l}=\left(\begin{array}{ccccccccc}
    \lambda-w_n^{l+m-1}&-w_n^{l+m-2}&-w_n^{l+m-3}&-w_n^{l+m-4}&\cdots&-w_n^{l+1}&-w_n^{l}&-v_n^{l} \\
    \lambda&-w_n^{l+m-2}&-w_n^{l+m-3}&-w_n^{l+m-4}&\cdots&-w_n^{l+1}&-w_n^{l}&-v_n^{l}\\
    \lambda&0 &-w_n^{l+m-3}&-w_n^{l+m-4}&\cdots&-w_n^{l+1}&-w_n^{l}&-v_n^{l}\\
    \vdots&\vdots&\vdots&\vdots&\ddots&\vdots&\vdots&\vdots\\
    \lambda&0&0&0&\cdots&-w_n^{l+1}&-w_n^{l}&-v_n^{l}\\
    \lambda&0&0&0&\cdots&0&-w_n^{l}&-v_n^{l}\\
    \lambda&0&0&0&\cdots&0&0&-v_n^{l}
  \end{array}
\right)_{(m+1)\times(m+1)}
\]
and
\[
M_{n}^{l}=\left(\begin{array}{cc}
    \begin{array}{cccccc}
      1&0&0&\cdots&0&\frac{1}{\lambda}(v_n^{l+1}-w_{n}^l)
    \end{array}
    &-\frac{v_n^l}{\lambda}\\
    I_m&
    \begin{array}{c}
      0\\
      0\\
      \vdots\\
      0
    \end{array}
  \end{array}
\right)_{(m+1)\times(m+1)},
\]
where $I_m$ denotes the unit matrix of order $m$.
The compatibility of the two linear systems in (\ref{dhrlv1-lax}) may be written as
\[
\begin{array}{rcl}
L_{n}^{l+1}M_{n}^{l}-M_{n+1}^{l}L_{n}^{l}=0 \; ,
\end{array}
\]
which leads to (\ref{dhRLV1}) and (\ref{dhRLV2}). Note that when $m=1$, this Lax pair reduces to
\[
L_{n}^{l}=\left(\begin{array}{cc}
\lambda-w_{n,l}&-v_{n,l}\\
\lambda&-v_{n,l}
\end{array}
\right),\ \ \
M_{n}^{l}=\left(\begin{array}{cc}
1+\frac{1}{\lambda}(v_{n,l+1}-w_{n,l})&-\frac{1}{\lambda}v_{n,l}\\
1&0
\end{array}
\right),
\]
which is in accordance with \cite{papageorgiou1995orthogonal,spicer2011higher}.

  Until now, we have indicated that the system (\ref{dhRLV1}) and (\ref{dhRLV2}) related to the QD-type dhLV equation can be derived by using the compatibility of the relations among the polynomials $P_n^l$.
  We next use the polynomials $Q_n^l$ to construct the dhQD algorithm
  \begin{eqnarray}
  \tilde w_n^{l+1}\tilde v_n^{l+m}=\tilde w_n^{l}\tilde v_{n+1}^{l} \;,\label{dhQD1}\\
  \tilde w_{n-1}^{l+1}+\tilde v_n^{l+m}=\tilde w_n^{l}+\tilde v_n^{l}.\label{dhQD2}
  \end{eqnarray}
 Similar to $P_n^l$, the crucial relations are
\begin{eqnarray*}
Q_{n}^{l+1}=\frac{1}{x}(Q_{n+1}^l+\tilde v_n^lQ_n^l) \; ,\\
Q_{n}^{l}=Q_{n}^{l+m}+\tilde w_{n-1}^lQ_{n-1}^{l+m} \; ,
\end{eqnarray*}
with
\[
\tilde v_n^l=\frac{\tau_{n+1}^{l+1}\tau_{n}^{l}}{\tau_{n}^{l+1}\tau_{n+1}^{l}} \; ,\ \ \ \
\tilde w_n^l=\frac{\tau_{n+2}^{l}\tau_{n}^{l+m}}{\tau_{n+1}^{l+m}\tau_{n+1}^{l}} \; ,
\]
which are obtained by using the two-row/column Sylvester identity. In addition, we can also get a Lax pair for the dhQD algorithm (\ref{dhQD1}) and (\ref{dhQD2}). Let
\[\Psi_n^l=(Q_n^{l+m},Q_n^{l+m-1},\cdots,Q_n^{l})^T,\]
 then
\[
\Psi_{n+1}^l=L_{n}^{l}\Psi_{n}^l \; ,\ \ \ \Psi_{n}^{l+1}=M_{n}^{l}\Psi_{n}^l \; ,
\]
with
\[
L_{n}^{l}=\left(\begin{array}{cccccccc}
    -\tilde w_n^l&0&0&\cdots&0&\lambda&-\tilde v_n^l \\
    \lambda&-\tilde v_n^{l+m-1}&0&\cdots&0&0&0\\
    \vdots&\lambda&-\tilde v_n^{l+m-2}&\cdots&0&0&0\\
    \vdots&\vdots&\vdots&\ddots&\vdots&\vdots&\vdots\\
    0&0&0&\cdots&-\tilde v_n^{l+2}&0&0\\
    0&0&0&\cdots&\lambda&-\tilde v_n^{l+1}&0\\
    0&0&0&\cdots&0&\lambda&-\tilde v_n^{l}
  \end{array}
\right)_{(m+1)\times(m+1)}
\]
and
\[
M_{n}^{l}=\left(\begin{array}{cc}
    \begin{array}{cccccc}
      -\frac{\tilde w_{n-1}^{l+1}}{\lambda}(1-\frac{\tilde v_{n-1}^{l+m}}{\tilde w_{n-1}^l})&0&0&\cdots&0&1
    \end{array}
    &-\frac{\tilde w_{n-1}^{l+1}\tilde v_{n-1}^{l+m}}{\lambda\tilde w_{n-1}^l}\\
    I_m&
    \begin{array}{c}
      0\\
      0\\
      \vdots\\
      0
    \end{array}
  \end{array}
\right)_{(m+1)\times(m+1)}.
\]
The compatibility condition $L_{n}^{l+1}M_{n}^{l}-M_{n+1}^{l}L_{n}^{l}=0$ is equivalent to the dhQD algorithm (\ref{dhQD1}) and (\ref{dhQD2}). Note also that this Lax pair reduces to the QD algorithm \cite{papageorgiou1995orthogonal,spicer2011higher} when $m=1$.

\section{The hHADT equation and the hQQD scheme from an orthogonal condition}
  In \cite{spicer2006orthogonal,spicer2011higher}, Spicer et al. have considered two-variable orthogonal polynomials, where the variables are restricted by the condition that they form the coordinates of an elliptic curve. More precisely, Spicer et al. have employed the Weierstrass elliptic curve
  \[
  y^2=x^3-ax-b
  \]
  and developed a sequence of elementary monomials
  \[
  e_0(x,y)=1,\ \  e_{2k}(x,y)=x^k,\ \ e_{2k+1}(x,y)=x^{k-1}y,\ \ k=1,2,\cdots
  \]
   associated with this curve as the basis of a space named $\mathcal{V}$. By defining an inner product $<,>$ on the space $\mathcal{V}$ and assuming that
  \[
  <xP,Q>=<P,xQ>
  \]
  for any two elements $P,Q\in\mathcal{V}$, a new class of two-variable adjacent orthogonal polynomials has been introduced:
  \[
  P_k^l(x,y)=
    \left|\begin{array}{ccccccccc}
    <e_l,e_0>&<e_l,e_2>&\cdots&<e_l,e_k>\\
    <e_{l+1},e_0>&<e_{l+1},e_2>&\cdots&<e_{l+1},e_k>\\
    \vdots&\vdots&\ddots&\vdots\\
    <e_{l+k-2},e_0>&<e_{l+k-2},e_2>&\cdots&<e_{l+k-2},e_k>\\
    e_0&e_2&\cdots&e_k
  \end{array}
\right|/\Delta_{k-1}^l, \ \ \ l=2,3,\cdots ,
  \]
  with
  \[
  \Delta_{k}^l=
    \left|\begin{array}{ccccccccc}
    <e_l,e_0>&<e_l,e_2>&\cdots&<e_l,e_k>\\
    <e_{l+1},e_0>&<e_{l+1},e_2>&\cdots&<e_{l+1},e_k>\\
    \vdots&\vdots&\ddots&\vdots\\
    <e_{l+k-1},e_0>&<e_{l+k-1},e_2>&\cdots&<e_{l+k-1},e_k>
  \end{array}
\right|, \ \ \ l=2,3,\cdots .
  \]
  These monomials and polynomials satisfy the orthogonal condition
  \[
  <e_l,P_k^l>=<e_{l+1},P_k^l>=\cdots=<e_{l+k-2},P_k^l>=0,\ \ \ l=2,3,\cdots
  \]

  The HADT equation (\ref{HADT}) and the QQD scheme (\ref{qqd}) are produced by the compatibility condition of
  several recurrence relations of $P_k^l$ and some auxiliary polynomials, with the help of
  the three-row/column Sylvester identity \cite{spicer2011higher,aitken1959determinants,brualdi1983determinantal}
  \begin{eqnarray*}
D D\left(\begin{array}{ccc}
i_1&i_2& n \\
j_1&j_2 & n \end{array}\right)&=&D\left(\begin{array}{cc}
i_1&i_2 \\
j_1&j_2 \end{array}\right)D\left(\begin{array}{c}
n  \\
n \end{array}\right)-D\left(\begin{array}{cc}
i_1&n  \\
j_1&j_2 \end{array}\right)D\left(\begin{array}{c}
i_2  \\
n \end{array}\right)\\
&&+D\left(\begin{array}{cc}
i_2&n  \\
j_1&j_2 \end{array}\right)D\left(\begin{array}{c}
i_1  \\
n \end{array}\right),
\end{eqnarray*}
where $i_1<i_2$ and $j_1<j_2$.

We next give generalizations to the HADT equation and the QQD scheme
by extending the orthogonal polynomials above. The methodology is similar to the generalization from
the QD algorithm to the dhQD. We call the generalizations the hungry forms of the HADT (hHADT) equation
and the QQD (hQQD) scheme, respectively.

 We begin with the adjacent elliptic orthogonal polynomials
\begin{equation*}
  P_k^l(x,y)=
    \left|\begin{array}{ccccccccc}
    <e_l,e_0>&<e_l,e_2>&\cdots&<e_l,e_k>\\
    <e_{l+m},e_0>&<e_{l+m},e_2>&\cdots&<e_{l+m},e_k>\\
    \vdots&\vdots&\ddots&\vdots\\
    <e_{l+(k-2)m},e_0>&<e_{l+(k-2)m},e_2>&\cdots&<e_{l+(k-2)m},e_k>\\
    e_0&e_2&\cdots&e_k
  \end{array}
\right|/\Delta_{k-1}^l \; ,
  \end{equation*}
  with
  \begin{equation*}
  \Delta_{k}^l=
    \left|\begin{array}{ccccccccc}
    <e_l,e_0>&<e_l,e_2>&\cdots&<e_l,e_k>\\
    <e_{l+m},e_0>&<e_{l+m},e_2>&\cdots&<e_{l+m},e_k>\\
    \vdots&\vdots&\ddots&\vdots\\
    <e_{l+(k-1)m},e_0>&<e_{l+(k-1)m},e_2>&\cdots&<e_{l+(k-1)m},e_k>
  \end{array}
\right|
  \end{equation*}
for a fixed positive integer $m$ and $l=2,3,\cdots$. Note that we
have omitted $m$ from $P_k^l(m,x,y)$ and $\Delta_{k}^l(m)$ for
simplicity. In addition, we introduce the polynomials
  \begin{equation*}
  Q_k^l(x,y)=
    \left|\begin{array}{ccccccccc}
    <e_l,e_0>&<e_l,e_2>&\cdots&<e_l,e_k>\\
    <e_{l+2m},e_0>&<e_{l+2m},e_2>&\cdots&<e_{l+2m},e_k>\\
    \vdots&\vdots&\ddots&\vdots\\
    <e_{l+(k-1)m},e_0>&<e_{l+(k-1)m},e_2>&\cdots&<e_{l+(k-1)m},e_k>\\
    e_0&e_2&\cdots&e_k
  \end{array}
\right|/\Theta_{k-1}^l \; ,
  \end{equation*}
  with
  \begin{equation*}
  \Theta_{k}^l=
    \left|\begin{array}{ccccccccc}
    <e_l,e_0>&<e_l,e_2>&\cdots&<e_l,e_k>\\
    <e_{l+2m},e_0>&<e_{l+2m},e_2>&\cdots&<e_{l+2m},e_k>\\
    \vdots&\vdots&\ddots&\vdots\\
    <e_{l+km},e_0>&<e_{l+km},e_2>&\cdots&<e_{l+km},e_k>
  \end{array}
\right|
  \end{equation*}
  for $l=2,3,\cdots.$

  It is easy to observe that
\[
  <e_l,P_k^l>=<e_{l+m},P_k^l>=\cdots=<e_{l+(k-2)m},P_k^l>=0,\ \ \ l=2,3,\cdots.
  \]
  Obviously, this orthogonal condition reduces to the one in \cite{spicer2006orthogonal} when $m=1$.

  Now we follow the same route in \cite{spicer2011higher} to deduce the new systems. Firstly, we list out
  the main relations in deriving the desired equations. Their proofs can be obtained by comparing
  with the corresponding formulae in \cite{spicer2011higher}. (The formula numbers in \cite{spicer2011higher}
  are given in the left parentheses.) We sketch the proofs in the appendix.
  \begin{eqnarray}
    (3.9)&&\rightarrow\ P_k^l=xP_{k-2}^{l+m+2}-V_{k-2}^lP_{k-1}^l+W_{k-2}^lP_{k-1}^{l+m}\label{relation_P1},\\
    &&{with}\ \ \  V_k^l=\frac{\Delta_{k}^{l}\Delta_{k}^{l+m+2}}{\Delta_{k+1}^{l}\Delta_{k-1}^{l+m+2}}, \ \ \ W_k^l=\frac{\Delta_{k}^{l+m}\Delta_{k}^{l+2}}{\Delta_{k+1}^{l}\Delta_{k-1}^{l+m+2}},\nonumber\\
  (3.12)&&\rightarrow\
    P_k^l=P_k^{l+m}+U_{k-1}^lP_{k-1}^{l+m}, \ \ {with}\ \ U_k^l=\frac{\Delta_{k+1}^{l}\Delta_{k-1}^{l+m}}{\Delta_{k}^{l}\Delta_{k}^{l+m}},\label{relation_P2}\\
  (3.14)&&\rightarrow\
    P_k^l=xQ_{k-2}^{l+2}-\frac{\Delta_{k-2}^{l}\Theta_{k-2}^{l+2}}{\Delta_{k-1}^{l}\Theta_{k-3}^{l+2}}P_{k-1}^l +\frac{\Delta_{k-2}^{l+2}\Theta_{k-2}^{l}}{\Delta_{k-1}^{l}\Theta_{k-3}^{l+2}}Q_{k-1}^l,\label{relation_PQ1}\\
    (3.15)&&\rightarrow\
    Q_k^l=P_k^{l+m}+\frac{\Delta_{k}^{l}\Delta_{k-2}^{l+2m}}{\Delta_{k-1}^{l+m}\Theta_{k-1}^{l}}P_{k-1}^{l+2m},\label{relation_PQ2}\\
    (3.16)&&\rightarrow\
    Q_k^l=P_k^l-\frac{\Delta_{k}^{l}\Theta_{k-2}^{l}}{\Delta_{k-1}^{l}\Theta_{k-1}^{l}}Q_{k-1}^l,\label{relation_PQ3}\\
    (3.17a)&&\rightarrow\ \Delta_k^l\Delta_{k-3}^{l+2m+2}=\Delta_{k-1}^{l}\Delta_{k-2}^{l+2m+2}-\Theta_{k-1}^{l}\Delta_{k-2}^{l+m+2} +\Delta_{k-1}^{l+m}\Theta_{k-2}^{l+2},\label{couple_system1}\\
    (3.17b)&&\rightarrow\ \Delta_{k}^{l}\Delta_{k-1}^{l+2m}=\Theta_{k}^{l}\Delta_{k-1}^{l+m}- \Delta_{k}^{l+m}\Theta_{k-1}^{l}.\label{couple_system2}
  \end{eqnarray}
\subsection{The hHADT equation}
  We move on to derive the hHADT equation by eliminating $\Theta_k^l$ from (\ref{couple_system1}) and (\ref{couple_system2}).
  These equations can be rearranged as
  \begin{eqnarray*}
    &&\frac{\Delta_k^l\Delta_{k-3}^{l+2m+2}}{\Delta_{k-2}^{l+m+2}\Delta_{k-1}^{l+m}}-\frac{\Delta_{k-1}^{l}\Delta_{k-2}^{l+2m+2}}{\Delta_{k-2}^{l+m+2}\Delta_{k-1}^{l+m}}=\frac{\Theta_{k-2}^{l+2}}{\Delta_{k-2}^{l+m+2}}-\frac{\Theta_{k-1}^{l}}{\Delta_{k-1}^{l+m}},\\
    &&\frac{\Delta_{k}^{l}\Delta_{k-1}^{l+2m}}{\Delta_{k-1}^{l+m}\Delta_{k}^{l+m}}=\frac{\Theta_{k}^{l}}{\Delta_{k}^{l+m}}- \frac{\Theta_{k-1}^{l}}{\Delta_{k-1}^{l+m}},
  \end{eqnarray*}
  which can be simplified to
  \begin{eqnarray*}
    &&X_k^l=\Gamma_k^l-\Gamma_{k-1}^l,\\
    &&Y_k^l=\Gamma_{k-2}^{l+2}-\Gamma_{k-1}^l ,
  \end{eqnarray*}
  where
  \begin{eqnarray*}
    &&X_k^l=\frac{\Delta_{k}^{l}\Delta_{k-1}^{l+2m}}{\Delta_{k-1}^{l+m}\Delta_{k}^{l+m}},\\
    &&Y_k^l=\frac{\Delta_k^l\Delta_{k-3}^{l+2m+2}}{\Delta_{k-2}^{l+m+2}\Delta_{k-1}^{l+m}}-\frac{\Delta_{k-1}^{l}\Delta_{k-2}^{l+2m+2}}{\Delta_{k-2}^{l+m+2}\Delta_{k-1}^{l+m}},\\
    &&\Gamma_k^l=\frac{\Theta_{k}^{l}}{\Delta_{k}^{l+m}}.
  \end{eqnarray*}
  Then we obtain
  \begin{eqnarray*}
    &&Y_{k+2}^l+X_{k+1}^l=\Gamma_{k}^{l+2}-\Gamma_{k}^{l},\\
    &&X_{k}^{l+2}+Y_{k+1}^l=\Gamma_{k}^{l+2}-\Gamma_{k}^{l}.
  \end{eqnarray*}
  Thus
  \[Y_{k+2}^l+X_{k+1}^l=X_{k}^{l+2}+Y_{k+1}^l\]
  is satisfied and can be equivalently expressed as
  \begin{equation}
  \begin{array}{l}
 \Delta_{k+1}^{l}(\Delta_{k-1}^{l+m+2}\Delta_{k}^{l+2m+2}\Delta_{k}^{l+m}-\Delta_{k-1}^{l+m+2}\Delta_{k}^{l+m+2}\Delta_{k}^{l+2m}+\Delta_{k-2}^{l+2m+2}\Delta_{k}^{l+m+2}\Delta_{k+1}^{l+m})\\
  =\Delta_{k-1}^{l+2m+2}(\Delta_{k}^{l+m}\Delta_{k-1}^{l+m+2}\Delta_{k+2}^{l}-\Delta_{k}^{l+m}\Delta_{k}^{l+2}\Delta_{k+1}^{l+m}+\Delta_{k}^{l+m+2}\Delta_{k+1}^{l+m}\Delta_{k}^{l}).
  \end{array}
  \end{equation}
  We call this equation the hungry form of the HADT (hHADT) equation.
  Obviously, it reduces to the HADT equation when $m=1$.

\subsection{A Lax pair for the system (\ref{couple_system1}) and (\ref{couple_system2}) }
Since the hHADT equation is obtained by eliminating $\Theta_k^l$
from (\ref{couple_system1}) and (\ref{couple_system2}), we may
regard (\ref{couple_system1}) and (\ref{couple_system2}) as a
coupled system by introducing an auxiliary variables $\Theta_k^l$ in
the hHADT equation. Now we present a Lax pair for this coupled system.

We let
{\small
$$\Psi_n^l=(Q_n^{l+2m+2},\cdots,Q_n^{l+m+1},P_{n+1}^{l+2m+2},\cdots,P_{n+1}^{l+m+1},Q_{n+1}^{l+m+2},\cdots,Q_{n+1}^{l+1},P_{n+2}^{l+m+2},\cdots,P_{n+2}^{l+1},)^T.$$
}
Then
\[
\Psi_{n+1}^l=L_{n}^{l}\Psi_{n}^l,\ \ \ \Psi_{n}^{l+1}=M_{n}^{l}\Psi_{n}^l,
\]
where $L_{n}^{l}$ and $M_{n}^{l}$ are respectively block matrices
\begin{equation}
L_{n}^{l}=\left(
\begin{array}{cccc}
  A_{11}&I_{m+2}&0&0\\
  0&A_{22}&0&I_{m+2}\\
  0&0&A_{33}&I_{m+2}\\
  0&0&A_{43}&A_{44}
\end{array}\right)
+\lambda \left(\begin{array}{cccc}
  0&0&0&0\\
  0&0&0&0\\
  0&0&0&0\\
  B_{41}&B_{42}&B_{43}&0
\end{array}\right) \; \label{hhadt-lax1},
\end{equation}
and
\begin{equation}
M_{n}^{l}=\left(
\begin{array}{cccc}
  C_{11}&C_{12}&C_{13}&0\\
  0&C_{22}&C_{23}&0\\
  C_{31}&C_{32}&C_{33}&0\\
  0&C_{42}&0&C_{44}
\end{array}\right)
+\frac{1}{\lambda} \left(\begin{array}{cccc}
  D_{11}&D_{12}&0&0\\
  0&0&0&0\\
  0&0&0&0\\
  0&0&0&0
\end{array}\right) \label{hhadt-lax2}.
\end{equation}
The explicit formulae will be presented in Appendix B because they are slightly cumbersome.

These relations can be verified by applying
(\ref{relation_P2})-(\ref{relation_PQ3}). Then the compatibility
condition $L_{n}^{l+1}M_{n}^{l}-M_{n+1}^{l}L_{n}^{l}=0$ yields the
system (\ref{couple_system1}) and (\ref{couple_system2}). Thus the
matrices $L$ and $M$ provide a Lax pair for the system
(\ref{couple_system1}) and (\ref{couple_system2}). We remark that,
when $m=1$, this Lax pair is different from that in
\cite{spicer2011higher}. In addition, we have given a unified
formula for all values of $m$.

\subsection{A Lax pair for the hQQD scheme and the hHADT equation}
  We now give a Lax pair that leads to the hQQD scheme directly and to the hHADT equation as well. Here we
  denote $U_k^l,V_k^l,W_k^l$ by $u_k^l,v_k^l,w_k^l$, respectively.

  Let
  \[
  \Psi_n^l=(P_{n}^{l+3m+2},\cdots,P_{n}^{l+2m+1},P_{n+1}^{l+2m+2},\cdots,P_{n+1}^{l+m+1},P_{n+2}^{l+m+2},\cdots,P_{n+2}^{l+1})^T,
  \]
  then
\[
\Psi_{n+1}^l=L_{n}^{l}\Psi_{n}^l,\ \ \ \Psi_{n}^{l+1}=M_{n}^{l}\Psi_{n}^l,
\]
where $L_{n}^{l}$ and $M_{n}^{l}$ are respectively block matrices
\begin{equation}
L_{n}^{l}=\left(
\begin{array}{ccc}
  A_{11}&I_{m+2}&0\\
  0&A_{22}&I_{m+2}\\
  0&A_{32}&A_{33}
\end{array}\right)
+\lambda \left(\begin{array}{ccc}
  0&0&0\\
  0&0&0\\
  B_{31}&B_{32}&0
\end{array}\right),\label{dhqqd-lax1}
\end{equation}

 and
 \begin{equation}
M_{n}^{l}=\left(
\begin{array}{ccc}
  C_{11}&0&0\\
  C_{21}&C_{22}&0\\
  0&C_{32}&C_{33}
\end{array}\right)
+\frac{1}{\lambda} \left(\begin{array}{ccc}
  D_{11}&D_{12}&0\\
  0&0&0\\
  0&0&0
\end{array}\right).\label{dhqqd-lax2}
\end{equation}
The explicit expressions are given in Appendix B.

From the compatibility condition
$L_{n}^{l+1}M_{n}^{l}-M_{n+1}^{l}L_{n}^{l}=0$, we get the hQQD
scheme
\begin{equation}
  \begin{array}{l}
  u_k^{l+3}w_k^{l+1}=u_{k+1}^{l+1}w_{k+1}^{l+1} ,\\
  u_k^{l+2+m}v_k^{l+m}=u_{k+1}^lv_{k+1}^l ,\\
  u_{k}^{l+2+m}+v_{k+1}^{l+m}+w_{k+1}^l=u_{k+2}^l+v_{k+1}^l+w_{k+1}^{l+m}.
  \end{array}\label{eq:hqqd}
  \end{equation}
  Thus $L$ and $M$ can be regarded as the Lax pair of the hQQD scheme.
 It is obvious that this scheme reduces to the QQD scheme when $m=1$.
 When one expresses $u,v,w$ in terms of $\Delta$, the third equation of (\ref{eq:hqqd}) yields the hHADT equation.
 Therefore $L$ and $M$ also provide a Lax pair for the hHADT equation. Note also
 that this Lax pair, in the case $m=1$, is different from that in \cite{spicer2011higher}.

 \section*{Acknowledgments}
This work was partially supported by the National Natural Science
Foundation of China (Grant no. 11331008, 11371251), the Knowledge Innovation
Program of LSEC, the Institute of Computational Mathematics, AMSS,
CAS, and the Hong Kong Research Grants Council (Grant no. GRF HKBU
202512).

 \appendix
\section{Determinant relations}
We sketch the derivations of
(\ref{relation_P1})-(\ref{couple_system2}) in this appendix.

In fact, (\ref{relation_P1}), (\ref{relation_PQ1}) and
(\ref{couple_system1}) are consequences of applying the
three-row/column Sylvester identity to (3.9), (B.9) and (3.17a),
respectively, in \cite{spicer2011higher}. We are left with proving
the other four relations.

We introduce the intermediate determinants
\begin{equation}
  T_k^l(x,y)=
    \left|\begin{array}{ccccccccc}
    <e_l,e_2>&<e_l,e_3>&\cdots&<e_l,e_{k+1}>\\
    <e_{l+m},e_2>&<e_{l+m},e_3>&\cdots&<e_{l+m},e_{k+1}>\\
    \vdots&\vdots&\ddots&\vdots\\
    <e_{l+(k-2)m},e_2>&<e_{l+(k-2)m},e_3>&\cdots&<e_{l+(k-2)m},e_{k+1}>\\
    e_2&e_3&\cdots&e_{k+1}
  \end{array}
\right|/\Pi_{k-1}^l \; ,
  \end{equation}
  with
  \begin{equation}
  \Pi_{k}^l=
    \left|\begin{array}{ccccccccc}
    <e_l,e_2>&<e_l,e_3>&\cdots&<e_l,e_{k+1}>\\
    <e_{l+m},e_2>&<e_{l+m},e_3>&\cdots&<e_{l+m},e_{k+1}>\\
    \vdots&\vdots&\ddots&\vdots\\
    <e_{l+(k-1)m},e_2>&<e_{l+(k-1)m},e_3>&\cdots&<e_{l+(k-1)m},e_{k+1}>
  \end{array}
\right|
  \end{equation}
for $l=2,3,\cdots.$

Using the two-row/column Sylvester identity as in {\textbf {Appendix
B}} of \cite{spicer2011higher}, we have
\begin{eqnarray}
&&(B.3a)\rightarrow\
  P_k^l=T_{k-1}^{l+m}-\frac{\Delta_{k-2}^{l+m}\Pi_{k-1}^l}{\Delta_{k-1}^{l}\Pi_{k-2}^{l+m}}P_{k-1}^{l+m},\label{relation_PT1}\\
  &&(B.3b)\rightarrow\ P_k^l=T_{k-1}^{l}-\frac{\Delta_{k-2}^{l}\Pi_{k-1}^l}{\Delta_{k-1}^{l}\Pi_{k-2}^{l}}P_{k-1}^{l},\\
  &&(B.8)\rightarrow \Delta_{k}^{l}\Pi_{k-2}^{l+m}=\Delta_{k-1}^{l}\Pi_{k-1}^{l+m} -\Delta_{k-1}^{l+m}\Pi_{k-1}^{l}.\label{jacobi_delta}
\end{eqnarray}
These three formulae lead to the derivation of (\ref{relation_P2}).

If we define $\Sigma_{k}^{l}$ by
\begin{equation}
  \Sigma_{k}^l=
    \left|\begin{array}{ccccccccc}
    <e_l,e_2>&<e_l,e_3>&\cdots&<e_l,e_{k+1}>\\
    <e_{l+2m},e_2>&<e_{l+2m},e_3>&\cdots&<e_{l+2m},e_{k+1}>\\
    \vdots&\vdots&\ddots&\vdots\\
    <e_{l+km},e_2>&<e_{l+km},e_3>&\cdots&<e_{l+km},e_{k+1}>
  \end{array}
\right|,\ \ \ l=2,3,\cdots,
  \end{equation}
  then
  \begin{eqnarray}
    &&(B.5a)\rightarrow \Delta_{k}^{l}\Pi_{k-2}^{l+2m}=\Pi_{k-1}^{l+m}\Theta_{k-1}^{l}-\Sigma_{k-1}^{l}\Delta_{k-1}^{l+m},\label{jacobi_deltapisitasigma}\\
  &&(B.5b)\rightarrow \Theta_{k}^{l}\Pi_{k-2}^{l+2m}=\Pi_{k-1}^{l+2m}\Theta_{k-1}^{l}-\Sigma_{k-1}^{l}\Delta_{k-1}^{l+2m}.
  \end{eqnarray}
  Eliminating $\Sigma$ from the two equations above, we get
  \begin{equation}
    (B.7)\rightarrow \Delta_{k-1}^{l+m}(\Pi_{k-1}^{l+2m}\Theta_{k-1}^{l}-\Theta_{k}^{l}\Pi_{k-2}^{l+2m})= \Delta_{k-1}^{l+2m}(\Pi_{k-1}^{l+m}\Theta_{k-1}^{l}-\Delta_{k}^{l}\Pi_{k-2}^{l+2m}).
  \end{equation}
  With the help of (\ref{jacobi_delta}), we obtain
  \begin{equation*}
    \Delta_{k-1}^{l+m}(\Pi_{k-1}^{l+2m}\Theta_{k-1}^{l}-\Theta_{k}^{l}\Delta_{k-2}^{l+2m})=
     \Theta_{k-1}^{l}(\Delta_{k-1}^{l+m}\Pi_{k-1}^{l+2m}-\Delta_{k}^{l+m}\Pi_{k-2}^{l+2m})
      -\Delta_{k-1}^{l+2m}\Delta_{k}^{l}\Pi_{k-2}^{l+2m},
    \end{equation*}
from which (\ref{couple_system2}) is derived after cancelling some terms.

 In addition, by employing the two-row/column Sylvester identity, we can also
 arrive at some intermediate relations for $P_k^l$ and $Q_k^l$:
  \begin{eqnarray}
    &&(B.10)\rightarrow\ Q_k^l=T_{k-1}^{l+2m}-\frac{\Delta_{k-2}^{l+2m}\Sigma_{k-1}^{l}}{\Pi_{k-2}^{l+2m}\Theta_{k-1}^{l}}P_{k-1}^{l+2m},\label{relation_PQT}\\
    &&(B.13a)\rightarrow\  Q_k^l=S_{k-1}^{l}-\frac{\Sigma_{k-1}^{l}\Theta_{k-2}^{l}}{\Sigma_{k-2}^{l}\Theta_{k-1}^{l}}Q_{k-1}^{l}, \label{relation_PQS1}  \\
    &&(B.13b)\rightarrow\ P_k^l=S_{k-1}^{l}-\frac{\Pi_{k-1}^{l}\Theta_{k-2}^{l}}{\Sigma_{k-2}^{l}\Delta_{k-1}^{l}}Q_{k-1}^{l},\label{relation_PQS2}\\
    &&(B.14)\rightarrow\ \Delta_{k}^{l}\Sigma_{k-2}^{l}=\Delta_{k-1}^{l}\Sigma_{k-1}^{l}- \Pi_{k-1}^{l}\Theta_{k-1}^{l},\label{relation_PQS3_sigma}
  \end{eqnarray}
  where
  \[
  \Sigma_{k}^l=
    \left|\begin{array}{ccccccccc}
    <e_l,e_2>&<e_l,e_3>&\cdots&<e_l,e_{k+1}>\\
    <e_{l+2m},e_2>&<e_{l+2m},e_3>&\cdots&<e_{l+2m},e_{k+1}>\\
    \vdots&\vdots&\ddots&\vdots\\
    <e_{l+(k-1)m},e_2>&<e_{l+(k-1)m},e_3>&\cdots&<e_{l+(k-1)m},e_{k+1}>\\
    e_2&e_3&\cdots&e_{k+1}
  \end{array}
\right|,\ \ \ l=2,3,\cdots. \\
  \]

From (\ref{relation_PT1}), (\ref{jacobi_deltapisitasigma}) and
(\ref{relation_PQT}) one can deduce (\ref{relation_PQ2}). Lastly,
equation (\ref{relation_PQ3}) can be proved by using
(\ref{relation_PQS1})-(\ref{relation_PQS3_sigma}).

\section{Explicit expressions of Lax pairs in Section 3}
\textbf{B.1.} The explicit expressions of the Lax pair (\ref{hhadt-lax1}) and (\ref{hhadt-lax2}) are given by
\[
\begin{array}{l}
  A_{11}=diag\{-\frac{\Delta_{n+1}^{l+2m+2}\Theta_{n-1}^{l+2m+2}}{\Delta_{n}^{l+2m+2}\Theta_{n}^{l+2m+2}},-\frac{\Delta_{n+1}^{l+2m+1}\Theta_{n-1}^{l+2m+1}}{\Delta_{n}^{l+2m+1}\Theta_{n}^{l+2m+1}},\cdots,-\frac{\Delta_{n+1}^{l+m+1}\Theta_{n-1}^{l+m+1}}{\Delta_{n}^{l+m+1}\Theta_{n}^{l+m+1}}\},\\
  A_{22}=diag\{-\frac{\Delta_{n+2}^{l+m+2}\Delta_{n}^{l+2m+2}}{\Delta_{n+1}^{l+m+2}\Delta_{n+1}^{l+2m+2}},-\frac{\Delta_{n+2}^{l+m+1}\Delta_{n}^{l+2m+1}}{\Delta_{n+1}^{l+m+1}\Delta_{n+1}^{l+2m+1}},\cdots,-\frac{\Delta_{n+2}^{l+1}\Delta_{n}^{l+m+1}}{\Delta_{n+1}^{l+1}\Delta_{n+1}^{l+m+1}}\},\\
  A_{33}=diag\{-\frac{\Delta_{n+2}^{l+m+2}\Theta_{n}^{l+m+2}}{\Delta_{n+1}^{l+m+2}\Theta_{n+1}^{l+m+2}},-\frac{\Delta_{n+2}^{l+m+1}\Theta_{n}^{l+m+1}}{\Delta_{n+1}^{l+m+1}\Theta_{n+1}^{l+m+1}},\cdots,-\frac{\Delta_{n+2}^{l+1}\Theta_{n}^{l+1}}{\Delta_{n+1}^{l+1}\Theta_{n+1}^{l+1}}\},\\
  A_{43}=diag\{-\frac{\Delta_{n+1}^{l+m+4}\Theta_{n}^{l+m+2}}{\Delta_{n+1}^{l+m+2}\Theta_{n}^{l+m+4}},-\frac{\Delta_{n+1}^{l+m+3}\Theta_{n}^{l+m+1}}{\Delta_{n+1}^{l+m+1}\Theta_{n}^{l+m+3}},\cdots,-\frac{\Delta_{n+1}^{l+3}\Theta_{n}^{l+1}}{\Delta_{n+1}^{l+1}\Theta_{n}^{l+3}}\},\\
  A_{44}=diag\{\frac{\Delta_{n+1}^{l+m+4}\Theta_{n+1}^{l+m+2}-\Delta_{n+1}^{l+m+2}\Theta_{n+1}^{l+m+4}}{\Delta_{n+2}^{l+m+2}\Theta_{n}^{l+m+4}},\frac{\Delta_{n+1}^{l+m+3}\Theta_{n+1}^{l+m+1}-\Delta_{n+1}^{l+m+1}\Theta_{n+1}^{l+m+3}}{\Delta_{n+2}^{l+m+1}\Theta_{n}^{l+m+3}},\cdots,\frac{\Delta_{n+1}^{l+3}\Theta_{n+1}^{l+1}-\Delta_{n+1}^{l+1}\Theta_{n+1}^{l+3}}{\Delta_{n+2}^{l+1}\Theta_{n}^{l+3}}\},\\
 \end{array}\]
\[
\begin{array}{l}
  B_{41}=\left(\begin{array}{ccccccc}
    0&\cdots&0&-\frac{\Delta_{n+1}^{l+m+4}\Theta_{n-1}^{l+m+4}}{\Delta_{n}^{l+m+4}\Theta_{n}^{l+m+4}}&0&0&0\\
    0&\cdots&0&0&-\frac{\Delta_{n+1}^{l+m+3}\Theta_{n-1}^{l+m+3}}{\Delta_{n}^{l+m+3}\Theta_{n}^{l+m+3}}&0&0\\
    0&\cdots&0&0&0&0&0\\
    \vdots&\ddots&\vdots&\vdots&\vdots&\vdots&\vdots\\
    0&\cdots&0&0&0&0&0
  \end{array}\right)_{(m+2)\times(m+2)},\\
  \end{array}\]
\[
\begin{array}{l}
  B_{42}=\left(\begin{array}{ccccccc}
    0&\cdots&0&1&0&0&0\\
    0&\cdots&0&0&1&0&0\\
    0&\cdots&0&0&0&0&0\\
    \vdots&\ddots&\vdots&\vdots&\vdots&\vdots&\vdots\\
    0&\cdots&0&0&0&0&0
  \end{array}\right)_{(m+2)\times(m+2)},\ \ \
  B_{43}=\left(\begin{array}{ccccccc}
  \begin{array}{ccc}
    0&\cdots&0\\
    0&\cdots&0
  \end{array}
  &\begin{array}{cc}
    0&0\\
    0&0
  \end{array}\\
    I_m&
    \begin{array}{cc}
      0&0\\
      \vdots&\vdots\\
      0&0
    \end{array}
  \end{array}\right)_{(m+2)\times(m+2)} ,\\
\end{array}\]

\[
\begin{array}{l}
  C_{11}=C_{33}=\left(\begin{array}{ccccccc}
    \begin{array}{ccc}
      0&\cdots&0
    \end{array}&0\\
    I_{m+1}&
    \begin{array}{c}
      0\\
      \vdots\\
      0
    \end{array}
  \end{array}\right)_{(m+2)\times(m+2)},\\
  \end{array}
  \]
  \[
\begin{array}{l}
  C_{12}=\left(\begin{array}{ccccccc}
    0&\cdots&0&\frac{\Delta_{n}^{l+m+3}\Delta_{n}^{l+2m+3}}{\Delta_{n+1}^{l+m+3}\Delta_{n-1}^{l+2m+3}}\left(1-\frac{\Delta_{n}^{l+m+3}\Theta_{n+1}^{l+3}}{\Delta_{n+1}^{l+3}\Delta_{n}^{l+2m+3}}\right)&0&0\\
    0&\cdots&0&0&0&0\\
    \vdots&\ddots&\vdots&\vdots&\vdots&\vdots\\
    0&\cdots&0&0&0&0
  \end{array}\right)_{(m+2)\times(m+2)},\\
  \end{array}\]
\[
\begin{array}{l}
  C_{13}=\left(\begin{array}{ccccccc}
    0&\cdots&0&\frac{\Delta_{n}^{l+m+3}\Theta_{n}^{l+3}}{\Delta_{n+1}^{l+3}\Delta_{n-1}^{l+2m+3}}&0&0\\
    0&\cdots&0&0&0&0\\
    \vdots&\ddots&\vdots&\vdots&\vdots&\vdots\\
    0&\cdots&0&0&0&0
  \end{array}\right)_{(m+2)\times(m+2)},
  \end{array}\]
\[\begin{array}{l}
  C_{22}=\left(\begin{array}{ccccccc}
    \begin{array}{ccccc}
      0&\cdots&0&\frac{\Delta_{n}^{l+m+3}\Theta_{n+1}^{l+3}}{\Delta_{n+1}^{l+3}\Delta_{n}^{l+2m+3}}&0
    \end{array}&0\\
    I_{m+1}&
    \begin{array}{c}
      0\\
      \vdots\\
      0
    \end{array}
  \end{array}\right)_{(m+2)\times(m+2)},\\
  \end{array}\]
\[
\begin{array}{l}
  C_{23}=\left(\begin{array}{ccccccc}
    0&\cdots&0&-\frac{\Delta_{n+1}^{l+m+3}\Theta_{n}^{l+3}}{\Delta_{n+1}^{l+3}\Delta_{n}^{l+2m+3}}&0&0\\
    0&\cdots&0&0&0&0\\
    \vdots&\ddots&\vdots&\vdots&\vdots&\vdots\\
    0&\cdots&0&0&0&0
  \end{array}\right)_{(m+2)\times(m+2)},\\
  \end{array}\]
\[
\begin{array}{l}
  C_{31}=\left(\begin{array}{ccccccc}
    0&\cdots&0&-\frac{\Delta_{n+1}^{l+m+3}\Theta_{n-1}^{l+m+3}}{\Delta_{n}^{l+m+3}\Theta_{n}^{l+m+3}}&0&0\\
    0&\cdots&0&0&0&0\\
    \vdots&\ddots&\vdots&\vdots&\vdots&\vdots\\
    0&\cdots&0&0&0&0
  \end{array}\right)_{(m+2)\times(m+2)},\\
   \end{array}\]
\[\begin{array}{l}
  C_{32}=\left(\begin{array}{ccccccc}
    0&\cdots&0&1&0&0\\
    0&\cdots&0&0&0&0\\
    \vdots&\ddots&\vdots&\vdots&\vdots&\vdots\\
    0&\cdots&0&0&0&0
  \end{array}\right)_{(m+2)\times(m+2)},\\
  \end{array}\]
\[\begin{array}{l}
  C_{42}=\left(\begin{array}{ccccccc}
    0&\cdots&0&-\frac{\Delta_{n+2}^{l+3}\Delta_{n}^{l+m+3}}{\Delta_{n+1}^{l+3}\Delta_{n+1}^{l+m+3}}&0&0\\
    0&\cdots&0&0&0&0\\
    \vdots&\ddots&\vdots&\vdots&\vdots&\vdots\\
    0&\cdots&0&0&0&0
  \end{array}\right)_{(m+2)\times(m+2)},\\
  \end{array}\]
\[
\begin{array}{l}
  C_{44}=\left(\begin{array}{ccccccc}
    \begin{array}{ccccc}
      0&\cdots&0&1&0
    \end{array}&0\\
    I_{m+1}&
    \begin{array}{c}
      0\\
      \vdots\\
      0
    \end{array}
  \end{array}\right)_{(m+2)\times(m+2)},
\end{array}\]
\[
\begin{array}{l}
  D_{11}=\left(\begin{array}{ccccccc}
    0&\frac{\Delta_{n}^{l+2m+3}\Theta_{n-1}^{l+2m+1}}{\Delta_{n}^{l+2m+1}\Theta_{n-1}^{l+2m+3}}&0&\cdots&0\\
    0&0&0&\cdots&0\\
    \vdots&\vdots&\vdots&\ddots&\vdots\\
    0&0&0&\cdots&0
    \end{array}\right)_{(m+2)\times(m+2)},\\
    \end{array}\]
\[
\begin{array}{l}
  D_{12}=\left(\begin{array}{ccccccc}
    0&-\frac{\Delta_{n}^{l+2m+3}\Theta_{n-2}^{l+2m+3}}{\Delta_{n-1}^{l+2m+3}\Theta_{n-1}^{l+2m+3}}+\frac{\Delta_{n}^{l+2m+3}\Delta_{n}^{l+m+1}}{\Delta_{n-1}^{l+2m+3}\Delta_{n+1}^{l+m+1}}&0&\cdots&0&-\frac{\Delta_{n}^{l+2m+3}\Delta_{n}^{l+m+1}}{\Delta_{n-1}^{l+2m+3}\Delta_{n+1}^{l+m+1}}\\
    0&0&0&\cdots&0&0\\
    \vdots&\vdots&\vdots&\ddots&\vdots&\vdots\\
    0&0&0&\cdots&0&0
    \end{array}\right)_{(m+2)\times(m+2)}.\\
    \end{array}\]
\textbf{B.2.} The explicit expressions of the Lax pair (\ref{dhqqd-lax1}) and (\ref{dhqqd-lax2}) are given by
\[
\begin{array}{l}
  A_{11}=diag\{-u_{n}^{l+2m+2},-u_{n}^{l+2m+1},\cdots,-u_{n}^{l+m+1}\},\\
  A_{22}=diag\{-u_{n+1}^{l+m+2},-u_{n+1}^{l+m+1},\cdots,-u_{n+1}^{l+1}\},\\
  A_{32}=diag\{-w_{n+1}^{l+m+2}u_{n+1}^{l+m+2},-w_{n+1}^{l+m+1}u_{n+1}^{l+m+1},\cdots,-w_{n+1}^{l+1}u_{n+1}^{l+1}\},\\
  A_{33}=diag\{w_{n+1}^{l+m+2}-v_{n+1}^{l+m+2},w_{n+1}^{l+m+1}-v_{n+1}^{l+m+1},\cdots,w_{n+1}^{l+1}-v_{n+1}^{l+1}\},
\end{array}
\]
\[\begin{array}{l}
 B_{31}=\left(\begin{array}{ccccccc}
    \begin{array}{ccccc}
  0&\cdots&0&-u_n^{l+m+4}&0\\
  0&\cdots&0&0&-u_n^{l+m+3}
  \end{array}&
  \begin{array}{cc}
    0&0\\
    0&0\\
    \end{array}\\
    I_m&
    \begin{array}{cc}
      0&0\\
      \vdots&\vdots\\
      0&0
    \end{array}
  \end{array}\right)_{(m+2)\times(m+2)},\\
  \end{array}\]
\[
\begin{array}{l}
  B_{32}=\left(\begin{array}{ccccccc}
  \begin{array}{ccccc}
  0&\cdots&0&1&0\\
  0&\cdots&0&0&1
  \end{array}&
  \begin{array}{cc}
    0&0\\
    0&0\\
    \end{array}\\
    I_m&
    \begin{array}{cc}
      0&0\\
      \vdots&\vdots\\
      0&0
    \end{array}
  \end{array}\right)_{(m+2)\times(m+2)},
 \end{array} \]
\[
\begin{array}{l}
  C_{11}=C_{22}=C_{33}=\left(\begin{array}{ccccccc}
    \begin{array}{ccccc}
      0&\cdots&0&1&0
    \end{array}&0\\
    I_{m+1}&
    \begin{array}{c}
      0\\
      \vdots\\
      0
    \end{array}
  \end{array}\right)_{(m+2)\times(m+2)},\\
  \end{array}\]
\[
\begin{array}{l}
  C_{21}=\left(\begin{array}{ccccccc}
    0&\cdots&0&-u_n^{l+m+3}&0&0\\
    0&\cdots&0&0&0&0\\
    \vdots&\ddots&\vdots&\vdots&\vdots&\vdots\\
    0&\cdots&0&0&0&0
  \end{array}\right)_{(m+2)\times(m+2)},\\
  \end{array}\]
\[
\begin{array}{l}
  C_{32}=\left(\begin{array}{ccccccc}
    0&\cdots&0&-u_{n+1}^{l+3}&0&0\\
    0&\cdots&0&0&0&0\\
    \vdots&\ddots&\vdots&\vdots&\vdots&\vdots\\
    0&\cdots&0&0&0&0
  \end{array}\right)_{(m+2)\times(m+2)},
  \end{array}\]
\[
  \begin{array}{l}
  D_{11}=\left(\begin{array}{ccccccc}
    0&u_{n-1}^{l+2m+3}w_{n-1}^{l+2m+1}&0&\cdots&0&-u_{n-1}^{l+2m+3}v_{n-1}^{l+2m+1}\\
    0&0&0&\cdots&0&0\\
    \vdots&\vdots&\vdots&\ddots&\vdots&\vdots\\
    0&0&0&\cdots&0&0
    \end{array}\right)_{(m+2)\times(m+2)},\\
    \end{array}\]
\[
\begin{array}{l}
    D_{12}=\left(\begin{array}{ccccccc}
    0&-u_{n-1}^{l+2m+3}&0&\cdots&0\\
    0&0&0&\cdots&0\\
    \vdots&\vdots&\vdots&\ddots&\vdots\\
    0&0&0&\cdots&0
    \end{array}\right)_{(m+2)\times(m+2)}.
\end{array}\]
\section*{References}

\end{document}